 \documentclass[conference]{IEEEtran}
\usepackage{cite}
\usepackage{multirow}
\usepackage{graphicx}
\usepackage{amssymb,amsmath,times,wasysym}
\usepackage{pb-diagram}
\usepackage{psfrag,balance}
\usepackage{srcltx,verbatim,color} 
\usepackage[caption=false]{subfig} 
\newcounter{mycounter}
\usepackage[top=0.75in,bottom=1in,right=0.625 in,left=0.625 in]{geometry}

\newcounter{appendx}

\newcommand{\real}[1]{\Re\left({#1}\right)}

\renewcommand{\exp}[1]{\mathrm{e}^{#1}}

\newcommand{\Exp}[1]{\mathrm{exp}\left(#1\right)}

\newcommand{\err}[1]{\mathrm{erf}\left(#1\right)}

\newcommand{\diag}[1]{\mathrm{diag}\left(#1\right)}

\renewcommand{\log}[2][]{\mathrm{log}_{#1}\left(#2\right)}

\newcommand{\E}[2][]{\mathbb{E}_{#1}\!\left[{#2}\right]}

\newcommand{\Var}[1]{\mathbb{V} \mathrm{ar}\!\left[{#1}\right]}

\title{Performance Analysis of IRS-Assisted Cell-Free Communication\vspace{-5mm}}

 \author{\IEEEauthorblockN{Diluka Loku Galappaththige, Dhanushka Kudathanthirige, and Gayan Amarasuriya } 
	\IEEEauthorblockA{School of Electrical, Computer, and Biomedical Engineering, Southern Illinois University, Carbondale, IL, USA 62901\\Email: \{diluka.lg,  dhanushka.kudathanthirige, gayan.baduge\}@siu.edu \vspace{-5mm}}
}

\begin{document}
\bstctlcite{IEEEexample:BSTcontrol}
\vspace{0mm}
\maketitle
 
% abstract & keywords

\begin{abstract}
	In this paper, the feasibility of adopting an intelligent reflective surface (IRS) in a cell-free wireless communication system is studied. The received signal-to-noise ratio (SNR) for this IRS-enabled cell-free set-up is optimized by   adjusting  phase-shifts of the passive  reflective elements. Then, tight approximations for the  probability density function and the cumulative distribution function for    this  optimal SNR are derived for Rayleigh fading. To investigate the performance of this system model, tight bounds/approximations for the achievable rate and outage probability   are derived in closed form. The impact of discrete phase-shifts is modeled, and  the corresponding detrimental effects  are investigated by deriving an upper bound for the achievable rate in the presence of    phase-shift quantization errors. Monte-Carlo simulations  are used to validate our statistical characterization of the optimal SNR, and the corresponding    analysis is used to investigate the performance gains of the proposed system model.  We reveal that  IRS-assisted communications can boost the performance of  cell-free wireless architectures.

\end{abstract}

 \linespread{1.0}

% ===========================================================================
% sections
% ===========================================================================
\vspace{-0mm}
% ----------------------------------------------------------------------------
\section{Introduction}\label{sec:introduction}
% ----------------------------------------------------------------------------

Recently, wireless architectures based on the notion of cell-free have gained much interest \cite{Ngo2015,Ngo2017}. In a cell-free system set-up, the cell-boundaries can be relaxed, and thus, a vast number of access-points (APs) can be  spatially distributed to serve all users with a uniformly better  quality-of-service (QoS)  over a much larger geographical region \cite{Ngo2015,Ngo2017}.  Moreover, cell-free set-ups may render spectral/energy efficiency gains, mitigate  impediments caused by spatial-correlated fading in  compact/co-located antenna arrays, and circumvent shadow fading impairments    \cite{Ngo2015,Ngo2017}. Thus, cell-free architecture is a foundation for practically realizing extremely large antenna arrays for next-generation wireless standards. 

An intelligence reflective surface (IRS) consists of a large number of  passive reflectors, whose  reflective coefficients can be  adjusted to attain desired propagation effects for the  
impinging electromagnetic (EM) waves \cite{Renzo2019,Liaskos2018}. 
The feature of intelligently adjustable phase-shifts at an IRS can be used to boost the signal-to-noise ratio (SNR)  and  to mitigate co-channel interference at an intended destination through  constructive and destructive  signal combining, respectively \cite{Diluka2020}. This leads to the notion of  recycling of EM waves within a propagation medium, and thereby, spectral/energy efficiency gains and implementation cost reduction  can be realized as IRSs are made out of low-cost meta-atoms without active  radio-frequency (RF) chains/amplifiers \cite{Liaskos2018}.

\subsection{Our motivation} 
In this paper, we aim to investigate the feasibility of embedding an IRS within a cell-free set-up. Specifically,  our objective is to investigate the performance of an IRS-assisted cell-free set-up, and thereby, we explore the feasibility of jointly reaping the aforementioned   benefits of cell-free architectures and IRS-assisted wireless channels. Moreover, to the best of the authors knowledge, the fundamental performance metrics for an IRS-assisted cell-free set-up have not yet been reported in open literature. To this end, we aim to fill this important gap in IRS literature by presenting a performance analysis for an IRS-assisted cell-free set-up.

\subsection{A literature survey for cell-free architecture and performance analysis of IRS-assisted channels}\label{sec:literature}
In \cite{Ngo2015,Ngo2017}, the basic concept of cell-free architectures is investigated, and thereby, the performance metrics are compared against those of the  co-located antenna arrays. The analyses in \cite{Ngo2015,Ngo2017,Ngo2018} reveal that the cell-free  set-ups can outperform the co-located counterparts by serving users with a uniformly better QoS,  minimizing the impediments of spatial-correlation, and shortening  the end-to-end transmission distances to boost the overall energy/spectral efficiency  \cite{Ngo2015,Ngo2017}. 
Reference \cite{Nayebi2017} proposes max-min power optimization algorithms for cell-free massive  multiple-input multiple-output (MIMO). In \cite{Galappaththige2019},  the performance of cell-free massive MIMO with underlay spectrum sharing is investigated.

References \cite{Renzo2019,Liaskos2018} present core architectural design principles of IRSs for wireless communications.
Ray-tracing techniques are used in \cite{Ozdogan2020} to generate a novel path-loss model for IRS-assisted wireless channels.
In \cite{Wu2019}, joint optimization of precoder at the base-station (BS) and phase-shifts at the IRS is studied  through semi-definite relaxation and alternative optimization techniques.  
Reference \cite{Diluka2020} studies the fundamental performance limits of distributed IRS-assisted end-to-end channels  with Nakagami-$m$ fading channels. In \cite{Han2019}, by using the statistical channel state information (CSI), an optimal phase-shift design framework is developed to maximize the achievable rates of  IRS-assisted wireless channels. 
In \cite{Chen2019}, joint beamforming and reflecting coefficient designs are investigated for IRSs  to provision physical layer security. Reference \cite{Abeywickrama2020} proposes a practical IRS phase-shift adjustment model, and thereby, the achievable rate is maximized through jointly optimizing the transmit power and the BS beamformer by using alternative optimization techniques.

\subsection{Our contribution}\label{sec:Motivation}
In  above-referred prior research \cite{Wu2019,Diluka2020,Han2019,Abeywickrama2020,Chen2019} for IRS-assisted communications, a BS with either a single-antenna or  a co-located antenna  array is used.
%A key attribute to all aforementioned related prior research \cite{Wu2019,Diluka2020,Han2019,Abeywickrama2020,Chen2019} on the IRS-assisted communication is that the antenna arrays at the BSs are co-located. 
Having been inspired by this  gap in IRS/cell-free literature, in this paper, we investigate an IRS-assisted wireless channel embedded within a cell-free set-up over Rayleigh fading,  and thereby, we present  fundamental performance metrics. 
To this end, first, we invoke the central limit theorem (CLT)  to tightly approximate the end-to-end optimal SNR  to facilitate a mathematically tractable probabilistic characterization. Then, we derive  the probability density function (PDF) and the cumulative density function (CDF) of this approximated optimal SNR in closed-form. Thereby, we present a tight approximation to the outage probability. Moreover, we derive tight upper/lower bounds for the achievable rate. In particular, we investigate the   impediments of discrete phase-shifts in the presence of phase-shift quantization errors. Finally, we present a set of rigorous  numerical results to explore the performance gains of  the proposed system, and we validate the  accuracy of our analysis through Monte-Carlo simulations. From our numerical results, we observe that  by using an  IRS with  controllable phase-shift adjustments, the performance of cell-free wireless set-ups  can be enhanced. 

%{\color{red}To this end, we first statistically characterize the end-to-end optimal SNR by tightly approximating it via a mathematically tractable counterpart by using the central limit theorem (CLT) \cite{papoulis02}. Thereby, the probability density function (PDF) and the cumulative density function (CDF) of the approximated SNR is derived in closed-form. Moreover, the outage probability approximation and the average achievable rate lower/upper bounds are derived for the proposed IRS-assisted cell-free set-up.  A rigorous set of numerical results is presented to investigate the performance of the proposed system while, the accuracy of our analysis is validated via Monte-Carlo simulations. Our numerical results  show that the performance of the cell-free communication systems can be boosted via enabling controllable propagation environment through IRSs.}

\noindent
\textbf{Notation:} The transpose of vector $\mathbf y$ is denoted as $\mathbf y^{T}$. The expectation and variance of a random variable  $Y$ are represented by 
$\E[]{Y}$ and $\Var{Y}$, respectively.
$Y\sim\mathcal {CN}\left(\mu_Y,  \sigma_Y^{2}  \right) $ denotes that $Y$ is  complex-valued circularly symmetric Gaussian distributed with $\mu_Y$ mean and $\sigma_Y^{2}$ variance. Moreover, $C_n=\{0,1,\cdots,n\}$ and $C_n'= C_n/\{0\}$.

\vspace{0mm}
\begin{figure}[!t]\centering \vspace{0mm}
	\def\svgwidth{180pt} 
		\fontsize{8}{3}\selectfont 
	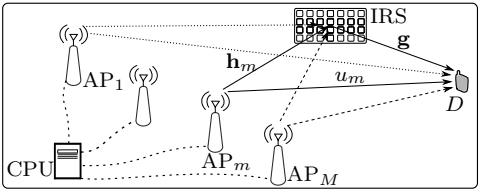 \vspace{-3mm}
	\caption{System model - IRS-aided cell-free communication set-up }\vspace{-6mm} \label{fig:system_model}
\end{figure}

%--------------------------------------------------------------------------------------
\section{System, Channel and Signal Models  }\label{sec:system_model}
%--------------------------------------------------------------------------------------

\subsection{System and channel model}\label{sec:system_and_channel}

We consider a cell-free communication set-up consisting of $M$   single-antenna APs ($\mathrm{AP}_m$ for $m = 1, \cdots, M$)  and a single-antenna destination $(D)$. An IRS having $N$ passive reflective elements is embedded within this cell-free set-up as shown in  Fig. \ref{fig:system_model}. 
%
%The system model of an IRS-enabled cell-free communication consists of $M$ number of single-antenna APs, a single-antenna destination $(D)$, and a single IRS having $N$ passive reflective elements as illustrated in Fig. \ref{fig:system_model}.
%We consider an IRS-enabled cell-free communication system  with $M$ number of single-antenna APs, a single-antenna destination $(D)$, and a single IRS having $N$ passive reflective elements (see Fig. \ref{fig:system_model}). 
For the sake of exposition, we denote the set of APs as $\mathcal M=\{1,\cdots, M\}$ and the set of reflective elements at the IRS as $\mathcal N=\{1,\cdots, N\}$.

The direct link between the $m$th AP and $D$ is represented by $u_m$, while  $h_{mn}$ denotes the channel between the $m$th AP and the $n$th reflective element of the IRS. Moreover, $g_n$ is used to represent the channel between the $n$th reflective element of the IRS and $D$.  We model the envelops of  all aforementioned channels to be independent Rayleigh distributed \cite{Ding2020}, and the corresponding polar-form of these channels is given by 
\begin{eqnarray}\label{eqn:channels}
	v = \lambda_v \exp{j\theta_{v}},
\end{eqnarray}
where $v\in \{u_m,h_{mn},g_{n}\}$ for $m\in \mathcal{M}$ and $n\in \mathcal{N}$. In \eqref{eqn:channels}, the envelop and the phase of $v$ are given by $\lambda_v$ and $\theta_{v}$, respectively. The  PDF of $\lambda_v$ is given by
\cite{papoulis02}
\begin{eqnarray}\label{eqn:chnl_pdf}
	f_{\lambda_v}(x) = \left({x}/{\xi_v}\right) \Exp{{- x^2}/{\left(2\xi_v\right)}},
\end{eqnarray} 
where $\xi_v = \zeta_v/2$ is the Rayleigh parameter, and $\zeta_v$ captures  the large-scale fading/path-loss of the channel $v$. Since all reflective elements are co-located within the IRS, it is assumed that all large-scale fading parameters are the same.

\subsection{Signal model}\label{sec:sgnl_modl}
The  signal transmitted by the $m$th AP reaches $D$ through the  direct and IRS-assisted reflected channels. Thus, we can write the signal received at $D$ as
\begin{eqnarray}\label{eqn:rx_signl}
r = \sqrt{P} \sum\nolimits_{m \in{\mathcal{M}}}  \left(u_m + \mathbf{g}^T  \mathbf{\Theta} \mathbf{h}_{m} \right) x + w,
\end{eqnarray}
where $x$ is the transmit signal from $S$ satisfying $\E{|x|^2} =1$,  $P$ is the  transmit power at each AP, and $w$ is an additive white Gaussian noise (AWGN) at $D$ with zero mean and variance of  $\sigma_{w}^2$ such that  $w\sim \mathcal{CN}(0,\sigma_{w}^2)$. In \eqref{eqn:rx_signl}, $\mathbf{h}_m = [h_{m1},\cdots, h_{mn}, \cdots, h_{mN}]^T\in \mathbb C^{N\times 1}$ is the channel vector between the $m$th AP and the IRS. Moreover,  $\mathbf{g}^T = [g_{1},\cdots, g_{n}, \cdots, g_{N}]\in \mathbb C^{1\times N}$ denotes the channel vector between the IRS and $D$. The diagonal matrix, $\mathbf \Theta = \diag{\beta_{1} \exp{j\theta_{1}}, \cdots, \beta_{n} \exp{j\theta_{n}}, \cdots, \beta_{N} \exp{j\theta_{N}}}\in \mathbb C^{N\times N}$, captures the reflective properties of the IRS through complex-valued reflection coefficients  $\beta_{n} \exp{j\theta_{n}}$ for $n\in \mathcal{N}$, where $\beta_{n}$ and $\theta_{n}$  are the magnitude of attenuation and phase-shift of the $n$th reflective element of the IRS, respectively. Thus, we can rewrite the received signal at $D$ in \eqref{eqn:rx_signl} as 
\begin{eqnarray}\label{eqn:rx_signl_rearng}
r = \sqrt{P}  \sum\nolimits_{m \in{\mathcal{M}}}  \left(u_m + \sum\nolimits_{n \in{\mathcal{N}}} \beta_{n} {g}_n  {h}_{mn} \exp{j\theta_{n}} \right) x + w.
\end{eqnarray}
Thereby, we derive the SNR at $D$ from \eqref{eqn:rx_signl_rearng} as
\begin{eqnarray}\label{eqn:snr}
\!\!\!\!\!\!\!\!\!\!\!\!\!\!\!\!\! \gamma &=&  \bar{\gamma} \left| \sum\nolimits_{m \in{\mathcal{M}}}  \left(u_m + \sum\nolimits_{n \in{\mathcal{N}}} \beta_{n} {g}_n  {h}_{mn} \exp{j\theta_{n}} \right) \right|^2 \nonumber \\
&=& \bar{\gamma} \left|  \sum\nolimits_{m \in{\mathcal{M}}} \!u_m  \!+\! \sum\nolimits_{n \in{\mathcal{N}}} \beta_{n} {g}_n \!  \left(\sum\nolimits_{m \in{\mathcal{M}}} {h}_{mn}\right) \exp{j\theta_{n}} \right|^2\!\!,
\end{eqnarray}
where the average transmit SNR is denoted by $\bar{\gamma}=P/\sigma_{w}^2$. Then, we define $u= \sum_{m \in{\mathcal{M}}} u_m$ and $h_n=\sum_{m \in{\mathcal{M}}} h_{mn}$. Since $u_m$ and $h_{mn}$ are independent complex Gaussian distributed for $m\in \mathcal{M}$ and $n\in \mathcal{N}$, the polar-form of $u$ and $h_n$ can be also expressed similar to \eqref{eqn:channels}, where $\lambda_{u}$ and $\lambda_{{h}_{n}}$ are the envelops of $u$ and $h_n$, respectively. Thus, $\lambda_{u}$ and $\lambda_{{h}_{n}}$ are  independent Rayleigh distributed with    parameters $\xi_u=\sum_{m \in{\mathcal{M}}} \zeta_{u_m}/2$ and $\xi_{h_n}=\sum_{m \in{\mathcal{M}}} \zeta_{h_{mn}}/2$, respectively. From \eqref{eqn:channels}, we can rewrite the SNR in \eqref{eqn:snr} in terms of the  channel phases as 
\begin{eqnarray}\label{eqn:snr_phase}
\gamma =  \bar{\gamma} \left| \lambda_{u} \exp{j\theta_{u}} + \sum\nolimits_{n \in{\mathcal{N}}} \beta_{n} \lambda_{{g}_n}  \lambda_{{h}_{n}} \exp{j\left(\theta_{n} + \theta_{{g}_n} + \theta_{{h}_{n}}\right)} \right|^2.
\end{eqnarray}
It can be seen from  \eqref{eqn:snr_phase} that the received SNR at $D$  can be maximized by smartly adjusting  the phase-shifts at each IRS reflecting elements $(\theta_{n})$. Thus, it enables a constructive addition of the received signals through the direct channels and IRS-aided reflected channels  \cite{Wu2019,Wu2020}. To this end, the optimal choice of $\theta_{n}$ is given by $\theta_{n}^* =\underset{-\pi\leq \theta_{n} \leq \pi}{ \mathrm{argmax}} \;{\gamma} = \theta_u - \left(\theta_{g_{n}} + \theta_{h_{n}}\right)$. Then,  we can derive the optimal SNR at $D$  as %\cite{Wu2019,Wu2020}
\begin{eqnarray}\label{eqn:snr_opt}
\gamma^* =  \bar{\gamma} \left| \lambda_{u} + \sum\nolimits_{n \in{\mathcal{N}}} \beta_{n} \lambda_{{g}_n}  \lambda_{{h}_{n}}  \right|^2.
\end{eqnarray}

%\noindent  \textit{ \textbf{Remark 1:} 
%Since in this work, we consider a single user case, the phase-shift optimization at the IRS is not necessarily needed for the rate performance analysis. On the other hand, in IRS-assisted cell-free massive MIMO with multiple user case, the phase-shift optimization must be done in order to analyze the performance of the system. To this end, the phase-shit optimization at the IRS for multiple user case is left as an open problem for the future research work.}

\section{Preliminaries}\label{sec:Preliminary_analysis}
In this section, we present a probabilistic  characterization of the optimal received SNR at $D$ in (\ref{eqn:snr_opt}). First, we denote the weighted sum of the product of  random variables in (\ref{eqn:snr_opt}) by  $Y=\sum_{n \in{\mathcal{N}}} \beta_{n} \lambda_{{g}_n}  \lambda_{{h}_{n}} $. Then,  we use the fact that $\lambda_{{g}_n}$ and  $\lambda_{{h}_{n}}$ for $n\in \mathcal{N}$  are independently distributed Rayleigh random variables to tightly approximate $Y$ through an one-sided  Gaussian distributed random variable $(\tilde{Y})$ by invoking the CLT \cite{papoulis02} as \cite{Diluka2020}
\begin{eqnarray}\label{eqn:pdf_Y}
\!\!\!\!\! f_Y(y) \approx 	f_{\tilde Y}(y) =
\frac{\psi}{\sqrt{2 \pi \sigma_{Y}^2}} \Exp{\!\frac{-(y-\mu_Y)^2}{2 \sigma_{Y}^2}\!}, \,  \text{for}  \,\, y\geq 0,
\end{eqnarray}
where  $\psi \triangleq 1/\mathcal{Q}\left(-\mu_Y/\sigma_{Y}\right)$ is a  normalization factor, which is used to ensure that  $\int_{-\infty}^{\infty} f_{\tilde Y}(x) dx=1$, and $\mathcal{Q}(\cdot)$ is the Gaussian-$\mathcal{Q}$ function \cite{papoulis02}. In (\ref{eqn:pdf_Y}), $\mu_Y$ and   $\sigma_{Y}^2$  are given by  
\begin{subequations}
	\begin{eqnarray} \label{eqn:mean_&_var}
	\mu_Y &=& \sum\nolimits_{n \in{\mathcal{N}}}  \pi \beta_{n} \left(\xi_{g_n} \xi_{h_n}\right)^{1/2}/2,\label{eqn:mean}\\
	\sigma_{Y}^2 &=& \sum\nolimits_{n \in{\mathcal{N}}}  \beta_{n}^2 \xi_{g_n} \xi_{h_n} \left(16-\pi^2\right)/4. \label{eqn:var}
	\end{eqnarray}  		 
\end{subequations}
Next, we derive a tight approximation for the PDF of $R=\lambda_{u}+Y$ as (see Appendix \ref{app:Appendix1})
\begin{eqnarray}\label{eqn:pdf_R}
\!\!\!\! f_R(x) \!&\approx&\! f_{\tilde R}(x)  \!=\! \sqrt{\pi} \rho \left(\frac{x-\mu_Y}{2\sigma_{Y}^2 \sqrt{a}}\right) \Exp{-\Delta \left(\frac{x-\mu_Y}{2\sigma_{Y}^2 \sqrt{a}}\right)^2} \nonumber \\
&&\!\!\!\!\!\!\!\! \!\!\!\! \times \left(\err{\frac{x-\mu_Y}{2\sigma_{Y}^2 \sqrt{a}}}+1\right) +\rho \Exp{- \left(\frac{x-\mu_Y}{2\sigma_{Y}^2 }\right)^2}\!, 
\end{eqnarray} 
 where $\err{x} = 2/\sqrt{\pi}\int_{0}^{x} \exp{-t} dt$ is the error function \cite[Eqn. 8.250.1]{Gradshteyn2007}. Here,  $a$, $\rho$,  and $\Delta$ are  given by
\begin{subequations}
\begin{eqnarray} \label{eqn:def_1}
\!\!\!\!\!\! a &=& {1}/{2\xi_u} + {1}/{2\sigma_{Y}^2},  \label{eqn:def_a} \qquad 
\rho = { \psi }\Big/\left({ 2a \xi_u  \sqrt{2\pi \sigma_{Y}^2}}\right), \label{eqn:def_digamma}\\
\!\!\!\!\!\! \Delta &=& \left(1-{1}/{2\sigma_{Y}^2} \right) 2\sigma_{Y}^2 a. \label{eqn:def_delta}	
\end{eqnarray}  		 
\end{subequations}
In particular, (\ref{eqn:pdf_R}) serves as the exact PDF of $\tilde{R}=\lambda_{u}+\tilde{Y}$, where $\tilde Y$ is the one-sided Gaussian approximated random variable for $Y$ in (\ref{eqn:snr_opt}).
Then, we derive an approximated PDF for $\gamma^* = \bar \gamma R^2 $ as
\begin{eqnarray}\label{eqn:pdf_gamma}
f_{\gamma^*}(y) &\approx& 
f_{\tilde R}\left(\sqrt{{y}/{\bar{\gamma}}}\right) \times {1}\big/{2 \sqrt{\bar{\gamma} y}}.
\end{eqnarray} 
Specifically, (\ref{eqn:pdf_gamma}) serves as the the exact PDF of $ \gamma^* \approx  \tilde{\gamma}^* = \bar \gamma \tilde {R}^2$.
From \eqref{eqn:pdf_R}, we derive the CDF of $\tilde{R}$ as (see Appendix \ref{app:Appendix2})
\begin{eqnarray}\label{eqn:cdf_R}
F_{\tilde{R}}(x) &=&  1 - \int_{x}^{\infty} f_{\tilde{R}}(u) du= 1 - \left(I_a+I_b\right) ,
\end{eqnarray}

\noindent where $I_a$ and $I_b$ are given by
\begin{subequations}
	\begin{eqnarray} \label{eqn:I_a_I_b}
	\!\!\!\!\!\!\! I_a &=& \frac{\lambda \exp{-\Delta d} \left(\err{d+1}\right)}{2\Delta}   + \frac{\lambda \left(1-\err{d \sqrt{\Delta+1}}\right)}{2\Delta \sqrt{\Delta+1}}, \label{eqn:I_a}\\
	\!\!\!\!\!\!\!I_b &=&  \sqrt{\frac{\pi \sigma_{Y}^2}{2} }\rho \left(1-\err{\sqrt{2 \sigma_{Y} a} d} \right), \label{eqn:I_b}	
	\end{eqnarray}  		 
\end{subequations}
where $\lambda = 2 \sigma_{Y}^2 \rho\sqrt{\pi a}$, 
$\rho$ is given in \eqref{eqn:def_digamma}, 
and $d=(x-\mu_Y)/(2\sigma_{Y}^2\sqrt{a})$. From \eqref{eqn:cdf_R}, we approximate the CDF of $\gamma^* = \bar \gamma R^2$  as
\begin{eqnarray}\label{eqn:cdf_SNR}
F_{\gamma^*}(y) &=&  \mathrm{Pr}\left(\gamma^*\leq y\right) \approx F_{\tilde R}\left(\sqrt{y/\bar{\gamma}} \right).
\end{eqnarray}

\begin{figure}[!t]\centering\vspace{-0mm}
	\includegraphics[width=0.38\textwidth]{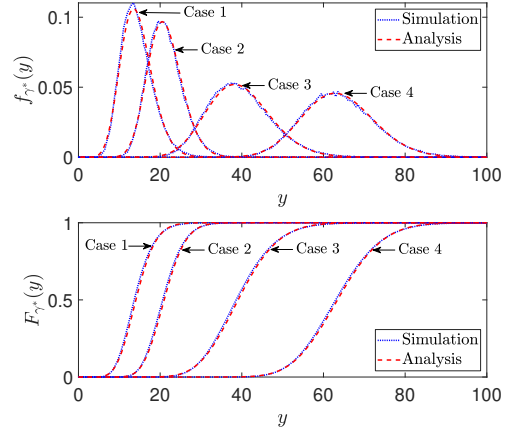}\vspace{-3mm}
	\caption{PDF and CDF of  SNR ($\gamma^*$) for $\bar{\gamma}= -10$dB. The combinations of $M$ and $N$ for Case 1 to Case 4 are set to $\{M=64, N=32\}$, $\{M=64, N=64\}$, $\{M=144, N=64\}$, and $\{M=64, N=128\}$. }
	\label{fig:PDF_CDF_M_N}\vspace{-6mm}
\end{figure}

\noindent  \textbf{\textit{Remark 1:}}  
	We plot the  approximated PDF and CDF of $\gamma^*$ by using the analysis in \eqref{eqn:pdf_gamma} and \eqref{eqn:cdf_SNR}, respectively, in Fig. \ref{fig:PDF_CDF_M_N}. Monte-Carlo simulations are also plotted in the same figure for various $M$ and $N$ to verify the accuracy of our approximations.
	From Fig. \ref{fig:PDF_CDF_M_N}, we   observe that  our analytical approximations for the  PDF \eqref{eqn:pdf_gamma} and CDF \eqref{eqn:cdf_SNR} of $\gamma^*$  are accurate even for moderately large values for  $M$ and $N$.

 %=====================================================================================
 \setcounter{mycounter}{\value{equation}}
 \begin{figure*}[!t] 
 	\addtocounter{equation}{1}
 	\vspace{-0mm}
 	\begin{small}
 		\begin{eqnarray} \label{eqn:rate_lb_sub}
 		\mathcal{R}_{lb} = \log[2]{1+ \frac{\bar{\gamma} \left(\xi_u +\sigma_{Y}^2 + 2\mu_{u} \mu_{Y} + \mu_{u}^2 + \mu_{Y}^2 \right)^3} {\sum\nolimits_{n\in C_4} \binom{4}{n} \left(2 \xi_u\right)^{n/2} \Gamma\left(n/2+1\right) \frac{\psi}{2\sqrt{\pi}} \sum\nolimits_{i \in C_n}  \binom{n}{i} \left( {2 \sigma_{Y}^2}\right)^{(n-i)/2}  \mu_{Y}^i I\left(n-i, \frac{-\mu_{Y}}{2 \sigma_{Y}^2}\right) }} 
 		\end{eqnarray}
 	\end{small} 
 	\vspace{-7mm}
% 	\hrulefill
 \end{figure*}
 \setcounter{equation}{\value{mycounter}}
 %%==================================================================================== 
 
  %=====================================================================================
 \setcounter{mycounter}{\value{equation}}
 \begin{figure*}[!t] 
 	\addtocounter{equation}{3}
 	\vspace{-0mm}
 	\begin{small}
 		\begin{eqnarray} \label{eqn:rate_ub_sub_q}
 		\hat{\mathcal{R}}_{ub} =   \log[2]{1+ \bar{\gamma} \left(\xi_u + {(\mu_{Y} \sin(\tau))}/{\tau} \left[2\mu_{u} + {(\mu_{Y} \sin(\tau))}/{\tau}\right] + {4 \sigma_{Y}^2}/({16-\pi^2}) \left[4- {\pi^2 \sin(\tau)^2}/{(4 \tau^2)}\right] \right)}
%\hat{\mathcal{R}}_{ub} =   \log[2]{1+ \bar{\gamma} \left(\xi_u + \frac{\mu_{Y} \sin(\tau)}{\tau} \left[2\mu_{u} + \frac{\mu_{Y} \sin(\tau)}{\tau}\right] + \left(\frac{4 \sigma_{Y}^2}{16-\pi^2} \right)\left[4- \frac{\pi^2 \sin(\tau)^2}{4 \tau^2}\right] \right)}
 		\end{eqnarray}
 	\end{small} 
 	\vspace{-7mm}
 	
 	\hrulefill
 	
 	\vspace{-7mm}
 \end{figure*}
 \setcounter{equation}{\value{mycounter}}
 %%==================================================================================== 

%-------------------------------------------------------------------------------------
\section{Performances Analysis}\label{sec:Performance_analysis}
%------------------------------------------------------------------------------------- 
\subsection{Outage probability }\label{sec:outage_prob}
An outage event occurs when the optimal received  SNR \eqref{eqn:snr_opt} falls below a threshold SNR ($\gamma_{th}$). To this end, we define the the outage probability of the proposed system model as 
$P_{out} = P_{r}\left(\gamma \leq \gamma_{th}\right)$. From \eqref{eqn:cdf_SNR}, we can compute the a tight approximation for the outage probability as $P_{out}   \approx F_{\gamma^*}(\gamma_{th})$.

\subsection{Average achievable rate }\label{sec:achvble_rate}
The average achievable rate of the proposed system can be defined as $
\mathcal{R} =   \E{\log[2]{1+\gamma^*}}$. 
The exact derivation of this expectation in $\mathcal{R}$   appears mathematically intractable. Thus,  we   resort to  tight upper/lower bounds for $\mathcal R$ as  $\mathcal{R}_{lb} \lesssim \mathcal{R} \lesssim \mathcal{R}_{ub}$ by invoking the Jensen's inequality \cite{Zhang2014}. 
Next, we derive  $\mathcal{R}_{ub}$  as (see Appendix \ref{app:Appendix3})
%
% where $\mathcal{R}_{lb}$ and $\mathcal{R}_{ub}$ are defined as
%\begin{subequations}
%\begin{eqnarray} 
%\mathcal{R}_{lb} &=& \log[2]{1 + \left(\E{1/\tilde{\gamma}^*}\right)^{-1}}, \label{eqn:rate_lb}\\
%\mathcal{R}_{ub} &=& \log[2]{1+ \E{\tilde {\gamma}^*}}. \label{eqn:rate_ub}	
%\end{eqnarray}  
%\end{subequations}
%
%
%The expectation term in \eqref{eqn:rate_ub} can be derived as 
%\begin{eqnarray}\label{eqn:E_gamma_ub}
%\E{\tilde{\gamma}^*} = \bar{\gamma} \left(\xi_u +\sigma_{Y}^2 + 2\mu_{u} \mu_{Y} + \mu_{u}^2 + \mu_{Y}^2 \right),
%\end{eqnarray} 
%\begin{proof1}
%	Refer Appendix \ref{app:Appendix3_1}.
%\end{proof1}
%
%\noindent where $\mu_{u} = \sqrt{\pi \xi_u/2}$. Thereby, we can derive the upper bound in \eqref{eqn:rate_ub}  as
\addtocounter{equation}{0}
\begin{eqnarray}\label{eqn:rate_ub_sub}
\mathcal{R}_{ub} =   \log[2]{1+ \bar{\gamma} \left(\xi_u +\sigma_{Y}^2 + 2\mu_{u} \mu_{Y} + \mu_{u}^2 + \mu_{Y}^2 \right)}.
\end{eqnarray}
We derive $\mathcal{R}_{lb}$ as given in \eqref{eqn:rate_lb_sub} at the top of the next page. 
\section{Impact of discrete phase-shift adjustments}\label{sec:Q_phase}
%-------------------------------------------------------------------------------------
Due to the hardware limitation, the adoption of continuous phase-shift adjustments for passive reflective elements at the IRS is practically challenging. Thus,  we investigate the feasibility of adopting discrete phase-shifts for the proposed set-up via phase-shift quantization. It is assumed that a limited number of discrete phase-shifts is available to select at the $n$th  reflector such that $\hat{\theta}_{n}^*= \pi \varsigma/2^{B-1}$, where $B$ denotes the number of quantization bits,    $\varsigma =\underset{q\in \{0,\pm 1, \cdots, \pm 2^{B-1} \} }{ \mathrm{argmin}} |{\theta}_{n}^* - \pi q/2^{B-1}| $, and $\theta_{n}^*$ is the optimal phase-shift in Section \ref{sec:sgnl_modl}.
Then, we can define the error of the continuous and quantized phase-shifts as $\varepsilon_n = {\theta}_{n}^*-\hat{\theta}_{n}^*$. For a large number of quantization levels, $\varepsilon_n$ can shown to be uniformly distributed as $\varepsilon_n \sim \mathcal{U} \left[-\tau,\tau \right) $ with $\tau=\pi/2^B$ \cite{Haykin2009}. The signal and   error $\varepsilon_n$ becomes uncorrelated for a high number of quantization levels \cite{Haykin2009}. Thus, the optimal SNR in \eqref{eqn:snr_opt} can be rewritten with discrete phase-shift as  
\begin{eqnarray}\label{eqn:snr_opt_q}
 	\addtocounter{equation}{1}
\!\!\!\!\!\! \hat{\gamma}^* \!=\!  \bar{\gamma} \left| \lambda_{u} \!+\! \sum_{n \in{\mathcal{N}}} \beta_{n} \lambda_{{g}_n}  \lambda_{{h}_{n}} \exp{j \varepsilon_n} \right|^2 \!=\!  \bar{\gamma} \left( (\lambda_{u} \!+\! Y_R)^2 \!+\!Y_I^2\right)\!,
\end{eqnarray}
where $Y_R = \sum_{n \in{\mathcal{N}}} \beta_{n} \lambda_{{g}_n}  \lambda_{{h}_{n}} \cos(\varepsilon_n)$ and $Y_I = \sum_{n \in{\mathcal{N}}} \beta_{n} \lambda_{{g}_n}  \lambda_{{h}_{n}} \sin(\varepsilon_n)$. By following steps similar to those in Appendix \ref{app:Appendix3},  an upper bound for the achievable rate with phase-shift quantization errors $(\hat{\mathcal{R}}_{ub})$ can be derived by using  \eqref{eqn:snr_opt_q} as shown in \eqref{eqn:rate_ub_sub_q}.
%\begin{eqnarray}\label{eqn:rate_ub_sub_q}
%\hat{\mathcal{R}}_{ub} =   \log[2]{1+ \bar{\gamma} \left(\xi_u + \frac{\mu_{Y} \sin(\tau)}{\tau} \left[2\mu_{u} + \frac{\mu_{Y} \sin(\tau)}{\tau}\right]  \right)}.
%\end{eqnarray}

%-------------------------------------------------------------------------------------
\section{Numerical Results}\label{sec:Numerical}
%-------------------------------------------------------------------------------------
The system parameters for  our simulations are given below: 
$\zeta_{v} = \left(d_0/d_{v}\right)^{\kappa} \times 10^{\varphi_{v}/10}$ is used to model large-scale fading, where  $v\in \{u_m,h_{mn},g_{n}\}$ for $m\in \mathcal{M}$ and $n\in \mathcal{N}$. The transmission distance between nodes is denoted by $d_v$,  $d_0=1$\,m is a  reference distance, the path-loss exponent is $\kappa=2.8$, and log-normal shadow fading is captured by $10^{\varphi_{v}/10}$  with $\varphi_{v} \sim (0,8)$ \cite{Marzetta2016_Book}. In our system topology, the IRS and $D$ are in positioned at fixed locations and  $250\,$m apart, while the APs  are uniformly distributed over an area of $1000 \times 1000$\,$\mathrm m^2$. The amplitudes of reflection coefficients are set to $\beta_{n}=0.9$ for $n \in \mathcal{N}$, which is a typical assumption for IRSs \cite{Wu2019,Wu2020}.

\begin{figure}[!t]\centering\vspace{-0mm}
	\includegraphics[width=0.38\textwidth]{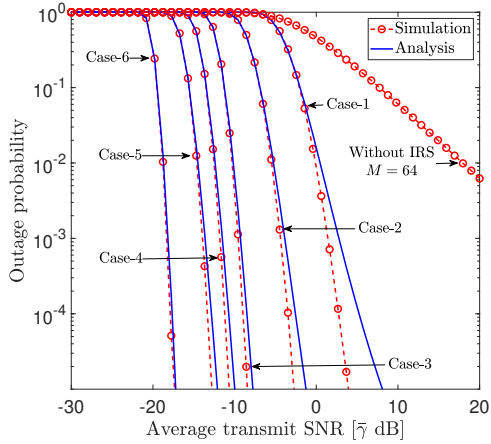}\vspace{-4mm}
	\caption{The outage probability for different $M$ and $N$ and $\gamma_{th}=0$ dB. The combinations of $M$ and $N$ for Case-1 to Case-6 are set to $\{M=36,N=16\}$, $\{M=36,N=32\}$, $\{M=16,N=64\}$, $\{M=36,N=64\}$, $\{M=64,N=64\}$, and $\{M=36,N=128\}$.}
	\label{fig:outage_M_N}\vspace{-5mm}
\end{figure}

In Fig. \ref{fig:outage_M_N}, we plot the outage probability as a function of the average transmit SNR ($\bar{\gamma}$) for different combinations of distributed APs $(M)$ and reflective elements  $(N)$ at the IRS. For comparison purposes, we also plot  the outage probability for the APs-to-$D$ direct transmission (without using an IRS) for $M=64$  in the same figure. 
%The analytical curves are plotted via our closed-form derivation in \eqref{eqn:out_prob}, while the accuracy of our analysis is validated by generating the exact outage curves via Monte-Carlo simulation.
We use our closed-form derivation in \eqref{eqn:cdf_SNR}  to plot the analytical outage probability approximations, and we plot the exact counterparts through Monte-Carlo simulation. The latter is used to verify the accuracy/tightness of our outage probability approximations.
According to Fig. \ref{fig:outage_M_N}, the tightness of our outage analysis  improves with as  $M$ or/and $N$ increase. 
The reason for this is that large $M$ or/and $N$ improves the accuracy of CLT.
Moreover, the outage probability can be reduced by either increasing $M$ or/and $N$. For example, at an average SNR of $-5\,$dB, the outage probability can be reduced by $99.9$\% by doubling $N$ from $16$ (Case-1) to $32$ (Case-2) while keeping $M=36$.
Moreover, by increasing  $M,N$ from $\{M=36,N=32\}$ in Case-2 to $\{M=64,N=64\}$ in Case-4, the average SNR required to achieve an outage probability of $10^{-3}$ can be reduced by $155.6\%$\,dB. 
From Fig. \ref{fig:outage_M_N}, we observe that the proposed IRS-aided cell-free  set-up  outperforms the APs-to-$D$ direct transmission. For instance,  the set-up without IRS needs an average transmit SNR of $18$\,dB to reach an outage probability of $10^{-2}$, which is about $177.6\%$  increase over the transmit SNR requirement for the Case-5 with the IRS-aided set-up for the same number of APs  $(M=64)$. Thus, the co-existence of IRSs within a cell-free set-up can be beneficial in reducing the system outage probability. 

\begin{figure}[!t]\centering\vspace{-0mm}
	\includegraphics[width=0.38\textwidth]{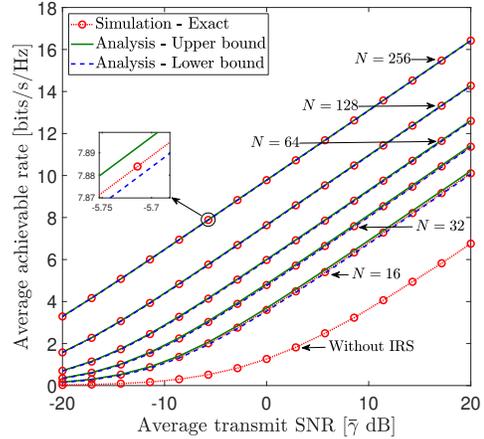}\vspace{-3mm}
	\caption{The average achievable   rate for $N \in \{16,32,64,128,256\}$ and $M=64$.}
	\label{fig:rate_N_0_16_32_64_128_256}\vspace{-6mm}
\end{figure}

In Fig. \ref{fig:rate_N_0_16_32_64_128_256}, we study the   average achievable rate of the proposed system as a function of  the average transmit SNR ($\bar{\gamma}$)  for $N \in \{16,32,64,128,256\}$.
%In Fig. \ref{fig:rate_N_0_16_32_64_128_256}, the average achievable rate is plotted as a function of average transmit SNR ($\bar{\gamma}$) for $N \in \{16,32,64,128,256\}$. 
We also compare the achievable   rates of APs-to-$D$ direct transmission and the IRS-aided transmission. 
The upper and lower bounds for the achievable rates are plotted by using our analysis in \eqref{eqn:rate_ub_sub} and \eqref{eqn:rate_lb_sub}, respectively. We again validate the accuracy of our analysis through Monte-Carlo simulations of the exact achievable rate. The tightness of our upper/lower rate bounds is clearly depicted in enlarged portion of Fig.  \ref{fig:rate_N_0_16_32_64_128_256}. We observe that the rate gains  can be achieved by increasing the  number of reflective elements in the IRS.  Fig.  \ref{fig:rate_N_0_16_32_64_128_256} also illustrates that an IRS can be embedded within a cell-free set-up to boost the achievable gains. For instance, an IRS with $N=16$  provides a rate gain of about $180$\,\% compared to the APs-to-$D$   transmission  without an IRS at an average transmit SNR of $0$\,dB.

%%==============Important======================================================================
%\begin{figure}[!t]\centering\vspace{-0mm}
%	\includegraphics[width=0.45\textwidth]{rate_M_16_36_64_144}\vspace{-3mm}
%	\caption{The average achievable   rate for $N = 64$ and $M= \in \{16,36,64,144\}$.}
%	\label{fig:rate_M_16_36_64_144}\vspace{-5mm}
%\end{figure}
%
%In Fig. \ref{fig:rate_M_16_36_64_144}, we study  the effect of the number of APs, and to this end,  the average achievable rates are plotted    by varying the number of APs as $M \in \{16,36,64,144\}$
%through \eqref{eqn:rate_ub_sub},  \eqref{eqn:rate_lb_sub} and Monte-Carlo simulations.
%We observe that by increasing the density of APs in a given geographical region for an IRS with a  fixed number of reflective elements ($N=64$), the achievable rate of the proposed system can be boosted. 
%For example, at an average transmit SNR of $-10$\,dB, the system with $M=36$ APs achieves a rate gain of $33.5\%$ compared to  the system with $M=16$ APs. 
%Moreover, in comparison to the system with $M=16$, the achievable rate  can be boosted by $92.5\%$ and $161.9\%$ for systems with $M=64$ and $M=144$, respectively.
%%===========================================================================================

\begin{figure}[!t]\centering\vspace{-0mm}
	\includegraphics[width=0.38\textwidth]{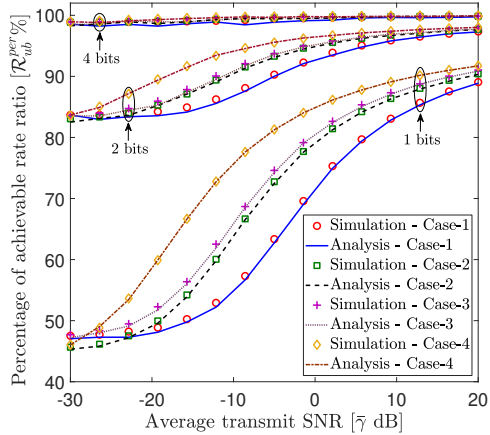}\vspace{-3mm}
	\caption{The impact of discrete phase-shifts with  phase-shift quantization on the average achievable rate for different $M$ and $N$. The combinations of $M$ and $N$ for Case-1 to Case-4 are set to $\{M=36,N=32\}$, $\{M=64,N=32\}$, $\{M=36,N=64\}$, and $\{M=64,N=64\}$.}
	\label{fig:rate_gain_M_N}\vspace{-6mm}
\end{figure}

In Fig. \ref{fig:rate_gain_M_N}, we investigate the impact of discrete phase-shifts and the  number of quantization bits ($B$) by plotting the  percentage rate ratio $(\mathcal{R}_{ub}^{per})$ against the average transmit SNR for different combinations of $M$ and $N$. The phase-shift quantization errors are uniformly distributed: $\mathcal{U} \left[-\pi/2^B, \pi/2^B \right)$.
The percentage rate ratio is defined as follows: $\mathcal{R}_{ub}^{per} = \hat{\mathcal{R}}_{ub}/\mathcal{R}_{ub} \times 100\%$, where $\hat{\mathcal{R}}_{ub}$ and $\mathcal{R}_{ub}$ are the upper bounds of the average achievable rate with and without phase-shift quantization errors given in \eqref{eqn:rate_ub_sub_q} and \eqref{eqn:rate_ub_sub}, respectively. Monte-Carlo simulation curves are also generated to validate our analysis. Fig. \ref{fig:rate_gain_M_N} shows  that the impact of phase-shift quantization errors vanishes when a higher $B$ is used. For instance, we can recover more than $98\%$ of the average rate when $4$ bit quantization is used at the IRS compared to the system with continuous phase-shift adjustments. As per Fig. \ref{fig:rate_gain_M_N}, $\mathcal{R}_{ub}^{per}$ improves in the high  SNR regime. For example, by varying $B$ as 1, 2, and 4 bits,  the average rate can be recovered more than $90\%$, $98\%$, and almost $100\%$, respectively, at a transmit SNR of $20\,$dB. Fig. \ref{fig:rate_gain_M_N} shows that a higher number of $M,N$ is also beneficial for recovering the achievable rate in the moderate-to-large transmit SNR regime.

% -------------------------------------------------------------------------------------
\section{Conclusion}\label{sec:conclusion}
% -------------------------------------------------------------------------------------
In this paper, the feasibility of adopting an IRS embedded within a   cell-free  set-up has been explored.
%The feasibility of IRS-assisted cell-free wireless communication network has been investigated. 
The optimal received SNR through multiple distributed APs with an IRS-aided channel has been  statistically characterized by deriving the tight PDF and CDF approximations. 
%the optimal SNR that can be achieved through the proposed system has been statistically characterized by deriving the tight approximations to the its exact PDF and CDF. 
This probabilistic SNR analysis has been  used to derive tight approximations/bounds for  the  outage probability and the average achievable rate in closed-form. 
%these approximated PDF/CDF have been used to derive the outage probability and the average achievable rate upper/lower bounds in closed-form. 
The impairments of discrete phase-shifts with equalization errors 
have been explored.
The accuracy  of our performance analysis of the proposed system set-up has been verified  by providing  Monte-Carlo simulations.
%The accuracy of our analysis has been validated by presenting a rigorous set of Monte-Carlo simulations.  
We observe from our numerical results that  IRS-aided cell-free system set-ups may be  used to reduce the outage probability and boost the achievable rates of next-generation wireless systems.

%======================================================================================= 
\appendices
%\linespread{1.0}

\section{The derivation of PDF of $\tilde{R}$ in \eqref{eqn:pdf_R} }\label{app:Appendix1}
By using the fact that $\lambda_u$ and $\tilde{Y}$ are independent random variables, we derive the PDF of $\tilde{R}$   as
\begin{eqnarray}\label{eqn:Apx_1_eqn_1}
f_{\tilde R}(x) &=& \int_{0}^{\infty} f_u(u) f_{\tilde Y}(x-u) du \\
&&\!\!\!\!\!\!\!\!\!\!\!\!\!\!\!\!\!\!\! 
= 2a \rho \exp{-\frac{(x-\mu_{Y})^2}{2\sigma_{Y}^2}} \int_{0}^{\infty} u \exp{-au^2+bu} du \nonumber \\
&&\!\!\!\!\!\!\!\!\!\!\!\!\!\!\!\!\!\!\!
= 2a \rho \exp{-\frac{(x-\mu_{Y})^2}{2\sigma_{Y}^2}} \exp{\frac{b^2}{4a}}  \int_{0}^{\infty} u \exp{-a \left(u-\frac{b}{2a}\right)^2} du \nonumber \\
&&\!\!\!\!\!\!\!\!\!\!\!\!\!\!\!\!\!\!\!
\stackrel{(a)}{=} 2a \rho \exp{-\frac{(x-\mu_{Y})^2}{2\sigma_{Y}^2}} \exp{\frac{b^2}{4a}} \!\! \left(\underbrace{\int_{-b/2a}^{\infty} \!\!\! t \exp{-at^2}dt  }_{I_1} + \frac{b}{2a} \underbrace{\int_{-b/2a}^{\infty}  \!\!\!\exp{-at^2}dt  }_{I_2}\right), \nonumber
\end{eqnarray} 
where $b = (x-\mu_{Y})/\sigma_{Y}^2$. The step $(a)$ is obtained by letting $t=u-b/2a$. Then, we can evaluate $I_1$ in \eqref{eqn:Apx_1_eqn_1} as 
\begin{eqnarray}\label{eqn:Apx_1_eqn_2}
\!\!\!\!\!\!\!\!  I_1 &=& \int_{-b/2a}^{\infty} \!\!\! t \exp{-at^2}dt \stackrel{(b)}{=} \left[-\exp{-at^2}/2a\right]_{-b/2a}^{\infty} =\exp{-b^2/2a},
\end{eqnarray} 
where the step $(b)$ is computed by using \cite[Eqn. 2.33.12]{Gradshteyn2007}.  Next, we evaluate $I_2$  as 
\begin{eqnarray}\label{eqn:Apx_1_eqn_3}
I_2 &=& \int_{-b/2a}^{\infty}  \exp{-at^2}dt \stackrel{(c)}{=} \left[\frac{\sqrt{\pi}\err{\sqrt{a}t}}{2\sqrt{a}}\right]_{-b/2a}^{\infty} \nonumber \\
&=& \frac{\sqrt{\pi}}{2\sqrt{a}} \left(1-\err{\frac{-b}{2\sqrt{a}}}\right),
\end{eqnarray}
where the step $(c)$ is due to \cite[Eqn. 2.33.16]{Gradshteyn2007}. We substitute \eqref{eqn:Apx_1_eqn_2} and \eqref{eqn:Apx_1_eqn_3} into \eqref{eqn:Apx_1_eqn_1} to obtain the PDF of $\tilde{R}$  in \eqref{eqn:pdf_R}.

\section{The derivation of CDF of $\tilde{R}$ in \eqref{eqn:cdf_R} }\label{app:Appendix2}
We substitute  \eqref{eqn:pdf_R} into  \eqref{eqn:cdf_R} to derive $I_a$  as 
\begin{eqnarray}\label{eqn:Apx_2_eqn_1}
I_a \!&=&\! \sqrt{\pi} \rho \!\!\! \int_{x}^{\infty} \!\!\! \left( \!\frac{u-\mu_Y}{2\sigma_{Y}^2 \sqrt{a}} \!\right) \! \exp{\!-\Delta \left(\!\frac{u-\mu_Y}{2\sigma_{Y}^2 \sqrt{a}}\!\right)^{\!2}} \!\! \left( \!\err{\frac{u-\mu_Y}{2\sigma_{Y}^2 \sqrt{a}}}\!+ \!1 \!\right) du \nonumber \\
&\stackrel{(d)}{=}& \lambda \int_{d}^{\infty} t \Exp{-\Delta t^2} \left(\err{t}+1\right) dt \nonumber \\
&\stackrel{(e)}{=}& \lambda \left[\frac{-\exp{-\Delta t^2} (\err{t}+1)}{2 \Delta}\right]_{d}^{\infty} + \lambda \int_{d}^{\infty} \frac{\exp{-t^2(\Delta+1)}}{2 \Delta} dt \nonumber \\
&\stackrel{(f)}{=}& \frac{\lambda \exp{-\Delta d} \left(\err{d+1}\right)}{2\Delta}   + \frac{\lambda \left(1-\err{d \sqrt{\Delta+1}}\right)}{2\Delta \sqrt{\Delta+1}},
\end{eqnarray}
where $\lambda=2 \sigma_{Y}^2 \rho\sqrt{\pi a}$ and $d=(x-\mu_Y)/(2\sigma_{Y}^2\sqrt{a})$. The step $(d)$ is  obtained by through $t=(u-\mu_{Y})/2\sigma_{Y}^2 \sqrt{a}$. The step $(e)$ is written by invoking  part-by-part integration, while the step $(f)$ is due to \cite[Eqn. 2.33.16]{Gradshteyn2007}. Next, we compute $I_b$ as    
\begin{eqnarray}\label{eqn:Apx_2_eqn_2}
I_b &=&  \rho  \int_{x}^{\infty} \exp{- \left(\frac{u-\mu_Y}{2\sigma_{Y}^2 }\right)^2} du \stackrel{(g)}{=} \sqrt{2 \sigma_{Y}^2} \rho \int_{\sqrt{2 \sigma_{Y} a} d}^{\infty} \exp{-t^2} dt \nonumber \\
&\stackrel{(h)}{=}&  \sqrt{\frac{\pi \sigma_{Y}^2}{2} }\rho \left(1-\err{\sqrt{2 \sigma_{Y} a} d} \right),
\end{eqnarray}
where the step $(g)$  is due to a changing of dummy variable as $t=(u-\mu_{Y})/(2\sigma_{Y}^2)$, and the step $(h)$ is resulted due to  \cite[Eqn. 2.33.16]{Gradshteyn2007}.

\section{The derivation of $\mathcal{R}_{lb}$ and $\mathcal{R}_{ub}$ in \eqref{eqn:rate_lb_sub} and \eqref{eqn:rate_ub_sub}}\label{app:Appendix3}
First,  by invoking Jensen's inequality, $\mathcal{R}_{lb}$ and $ \mathcal{R}_{ub}$  can be defined  as 
\begin{subequations}
	\begin{eqnarray} 
	\mathcal{R}_{lb} &=& \log[2]{1 + \left(\E{1/\tilde{\gamma}^*}\right)^{-1}}, \label{eqn:rate_lb}\\
	\mathcal{R}_{ub} &=& \log[2]{1+ \E{\tilde {\gamma}^*}}. \label{eqn:rate_ub}	
	\end{eqnarray}  
\end{subequations}
%\subsection{The derivation of expectation of $\gamma^*$ in \eqref{eqn:E_gamma_ub} }\label{app:Appendix3_1}
Then, we evaluate the expectation term in  \eqref{eqn:rate_ub}  as
\begin{eqnarray}\label{eqn:Apx_3_eqn_1}
\E{\tilde{\gamma}^*} &=& \E{\bar{\gamma} \tilde R^2} = \bar{\gamma}\E{(\lambda_u + \tilde Y)^2} \nonumber\\
&=& \bar{\gamma}  \sum_{n \in C_2} \!\! \binom{2}{n} \E{\lambda_u^{(2-n)}} \E{\tilde Y^n} \nonumber\\
&=& \bar{\gamma} \left(\xi_u + \mu_{u}^2 + \sigma_{Y}^2 + \mu_{Y}^2 + 2 \mu_{u}\mu_{Y}  \right), 
\end{eqnarray}
where $\mu_{u} = \sqrt{\pi \xi_u/2}$. Moreover, $\mu_{Y}$ and $\sigma_{Y}^2$ are given in \eqref{eqn:mean} and \eqref{eqn:var}, receptively. By substituting \eqref{eqn:Apx_3_eqn_1} into \eqref{eqn:rate_ub}, $\mathcal{R}_{ub}$ can be computed as \eqref{eqn:rate_ub_sub}.
Next, we can write the expectation term in \eqref{eqn:rate_lb} as  
\begin{eqnarray}\label{eqn:E_gamma_lb}
\E{1/\tilde{\gamma}^*} = {1}/{\E{\tilde\gamma^*}} + {\Var{\tilde\gamma^*}}/{\left(\E{\tilde\gamma^*}\right)^3},
\end{eqnarray}
where $\E{\tilde{\gamma}^*}$ is defined in \eqref{eqn:Apx_3_eqn_1} and $\Var{\tilde\gamma^*}= \bar{\gamma}^2\E{\tilde R^4} - \left(\E{\tilde\gamma^*}\right)^2$. Then, we can compute $\E{\tilde R^4}$ as follows: 
\begin{eqnarray}\label{eqn:Apx_3_eqn_2}
\!\!\!\!\!\!\!\! \E{\tilde R^4} &=& \E{(\lambda_u+ \tilde Y)^4}\! = \sum_{n \in C_4} \! \binom{4}{n} \E{\lambda_u^{(4-n)}} \E{\tilde Y^n},
\end{eqnarray}
where the $n$th moment of $\lambda_u^n$ is denoted by  $\E{\lambda_u^n}$. We compute $\E{\lambda_u^n}$ as
\begin{eqnarray}\label{eqn:Apx_3_eqn_3}
\E{\lambda_u^n} &=& \int_{0}^{\infty} x^n f_{u}(x) dx   =\int_{0}^{\infty} \frac{x^{n+1}}{\xi_u} \Exp{-\frac{x^2}{2\xi_u}} dx \nonumber \\
&\stackrel{(i)}{=}& \left(2\xi_u\right)^{n/2} \Gamma\left(n/2+1\right),  
\end{eqnarray}
where the step $(m)$ is evaluated from \cite[Eqn. 2.33.10]{Gradshteyn2007} and $\Gamma(t) =\int_{0}^{\infty} x^t \exp{-x} dx$ is the Gamma function \cite[Eqn. 8.310.1]{Gradshteyn2007}. Then, we  evaluate  $\E{\tilde Y^n}$  for $n \in C_4'$ as 
\begin{eqnarray}\label{eqn:Apx_3_eqn_4}
\E{\tilde Y^n} &=& \frac{\psi}{\sqrt{2\pi \sigma_{Y}^2}}  \int_{0}^{\infty} y^{n}  \exp{-\frac{(y-\mu_{Y})^2}{2\sigma_{Y}^2}} dy  \\
&\stackrel{(j)}{=}& \frac{\psi}{\sqrt{\pi }} \int_{{-\mu_{Y}}/{\sqrt{2\sigma_{Y}^2}}}^{\infty} \left(\sqrt{2\sigma_{Y}^2}t +\mu_{Y}\right)^n \exp{-t^2} dt \nonumber \\
%\end{eqnarray}
%\begin{eqnarray}
&\stackrel{(k)}{=}& \!\frac{\psi}{2\sqrt{\pi}} \sum\limits_{i\in C_n} \!\! \binom{n}{i} \left( {2 \sigma_{Y}^2}\right)^{\frac{n-i}{2}} \mu_{Y}^i I\!\left(n\!-\!i, \frac{-\mu_{Y}}{2 \sigma_{Y}^2}\right), \nonumber
\end{eqnarray}
where the step $(j) $ is due to a changing  of the dummy variable,  the step $(k)$ is obtained by expanding $\left( \!\sqrt{2\sigma_{Y}^2}t \!+\! \mu_{Y} \!\right)^{\!\!n}\!$ based on $n\!$ value. Moreover, $\!I(\!\cdot,\cdot\!)\!$ is given as
\begin{eqnarray}\label{eqn:I}
\!\!\!\!\!\!\!\!\! I\!\left(m,t\right) \!=\! \begin{cases}
(-1)^m \gamma\left(\frac{m+1}{2}, t^2\right) + \Gamma\left(\frac{m+1}{2}\right),  &\text{for} \,\, t \leq 0,  \\
\Gamma\left(\frac{m+1}{2}, t^2\right),& \text{otherwise},
\end{cases}
\end{eqnarray} 
where $\gamma(\lambda,x) =  \int_{0}^{x} \exp{-t} t^{\lambda-1}dt$ is the lower incomplete Gamma function \cite[Eqn. 8.350.1]{Gradshteyn2007}. Finally, $\mathcal{R}_{lb}$ is derived  as  \eqref{eqn:rate_lb_sub}.

  \linespread{1.0}
% ===========================================================================
% bibliography
% ===========================================================================
\bibliographystyle{IEEEtran}
\bibliography{IEEEabrv,References_1}%,References

\end{document}